\documentclass{ws-p8-50x6-00}
\newcommand{\fslash}[1]{\mbox{$\!\not\!#1$}}
\begin{document}

\title{Color Superconductivity at Moderate Density}
\author{Mei Huang$^{1}$, Pengfei Zhuang$^{1}$, Weiqin Chao$^{2,3}$ \\}
\address{$^1$ Physics Department, Tsinghua University, Beijing 100084, China \\
$^2$ CCAST, Beijing 100080, China\\
$^3$ IHEP, Chinese Academy of Sciences, Beijing 100039, China}  

\maketitle

\abstracts{The effect of color breaking on colored quarks' chiral condensates 
has been investigated at zero temperature and moderate baryon density. 
It is found that the influence of the diquark condensate on different
colored quarks is very small.}

The QCD phase structure along the baryon density direction
has attracted much attention recently since it was found that 
the gap of the color superconductivity can be of the order of 100 MeV 
due to the non-perturbative effects \cite{raja1}. From experimental point of 
view, the region of moderate density is of great interest,  since it is expected 
that the color superconductivity might be realized in heavy ion collisions 
and in the interior of the neutron stars. 
The model calculations in this region show a strong competition 
between chiral condensate and diquark condensate \cite{berges} 
\cite{carter} and \cite{kerbikov}, i.e., in the chiral limit, 
when diquark condensate appears, the chiral condensate disappears. 
Thus in the chiral limit, it is impossible to see the influence of the diquark condensate 
on different colored quarks. 
In the real physical case, the chiral condensate would not disappear 
entirely in the color superconducting phase, we will investigate the difference
between the quarks participating in the diquark condensate and the quark
which does not involve in the diquark condensate by calculating 
colored quarks' chiral condensate.   

The model we used is an extended Nambu--Jona-Lasinio model which is defined 
through the following Lagrangian density,
\begin{eqnarray}
\label{lagr}
{\cal L}={\bar q}(i {\fslash \partial}-m_0)q + 
   G_S[({\bar q}q)^2 + ({\bar q}i\gamma_5{\bf {\vec \tau}}q)^2 ] + 
  G_D[(i {\bar q}^C  \varepsilon  \epsilon^{\rho} \gamma_5 q )
   (i {\bar q} \varepsilon \epsilon^{\rho} \gamma_5 q^C)],
\end{eqnarray}
here $G_S$ and $G_D$ are independent effective coupling constants in the 
scalar quark-antiquark and scalar diquark channel.
$m_0$ is the current quark mass, assuming isospin degeneracy of $u$ and $d$ quarks. 

After bosonization, introducing the 8-component spinors 
$\Psi$ and $ \Psi_3 $ for the quarks participated and not participated
the diquark condensate, the partition function can be evaluated as
\begin{eqnarray}
\label{part}
{\cal Z} & = & N'{\rm exp} \{- \int_0^{\beta} d \tau \int d^3{\vec x} ~ 
[\frac{\sigma^2}{4 G_S}
      +\frac{\Delta^{+}\Delta^{-}}{4 G_D}] \} \nonumber \\
   & &  \int[d \Psi_3]{\rm exp}\{\frac{1}{2}\sum_{n,{\vec p}}  
 ~ {\bar \Psi}_3\frac{G_0^ {-1}}{T}\Psi_3 \}  
  \int[d \Psi]{\rm exp} \{\frac{1}{2}  \sum_{n,{\vec p}} ~
 {\bar \Psi}\frac{{\rm G}^{-1}}{T}  \Psi \},
\end{eqnarray}
where
\begin{equation}
{\rm G_0}^{-1} = 
    \left( 
          \begin{array}{cc}
            \left[ G_0^{+} \right]^{-1}  &  0 \\  
            0 &  \left[ G_0^{-} \right]^{-1} 
              \end{array}  
             \right),\ \ {\rm G}^{-1} = 
          \left( 
          \begin{array}{cc}
            \left[ G_0^{+} \right]^{-1}  &  \Delta^{-} \\  
            \Delta^{+}  &  \left[ G_0^{-} \right]^{-1} 
              \end{array}  
             \right),
\end{equation}
with $[G_0^{\pm}]^{-1}=
(p_0 \pm \mu) \gamma_0 -{\vec \gamma}\cdot {\vec p} -m, p_0=i \omega_n$ in 
the momentum space. 

Using the energy projector $\tilde \Lambda_{\pm}$ \cite{mei}, 
for the quark does not participate the diquark condensate, its propagator
can be expressed as   
\begin{eqnarray}
\label{mass0}
G_0^{\pm} =  \frac{\gamma_0\tilde \Lambda_{+}}{p_0+E_p^{\pm}} + 
\frac{\gamma_0\tilde\Lambda_{-}}{p_0-E_p^{\mp}}, 
\end{eqnarray}
while the propagators of the quarks participated the diquark condensate can be expressed as
\begin{eqnarray}
G^{\pm} & = & \frac{p_0-E_p^{\pm}}{p_0^2- {E_{\Delta}^{\pm}}^2}
\gamma_0\tilde\Lambda_{+}
  + \frac{p_0+E_p^{\mp}}{p_0^2-{E_{\Delta}^{\mp}}^2 }\gamma_0\tilde\Lambda_{-}, 
\end{eqnarray}
with $ E_p^{\pm} = E_p \pm \mu, \ \ {E_{\Delta}^{\pm}}^2={E_p^{\pm}}^2+\Delta^2$, and
$ E_p = \sqrt{p^2+m^2}$. 

In the mean-field approximation,  one can evaluated the thermodynamic potential as
\begin{eqnarray}
\Omega & = &  \frac{\sigma^2}{4G_S}+\frac{\Delta^2}{4G_D} 
-2N_f \int\frac{d^3 p}{(2\pi)^3} [ E_p + T{\rm ln}(1+e^{-\beta E_p^{+}}) 
 + T {\rm ln}(1+ e^{-\beta E_p^{-}})  \nonumber \\
&+ & E_{\Delta}^{+} 
+2T{\rm ln}(1+e^{-\beta E_{\Delta}^{+}}) + E_{\Delta}^{-} 
+2T {\rm ln}(1+ e^{-\beta E_{\Delta}^{-}}) ].
\end{eqnarray}
Extremizing $\Omega$ leads to the two coupled gap equations  
\begin{eqnarray}
\label{gap}
m & = &  m_0+\sigma, \ \  \sigma=-2G_S<{\bar q}q>, \nonumber \\
\Delta & = & -2G_D <{\bar q}^C \varepsilon \epsilon^{b}\gamma_5 q>,
\end{eqnarray}
where the chiral condensate is the sum
of different colored quarks' chiral condensates,i.e.,  
$<{\bar q}q>=2 <{\bar q_1}q^1> + <{\bar q_3}q^3> $, 
which can be evaluated by using the colored quark propagators 
\begin{eqnarray}
<{\bar q}_3 q^3> =  -iT \sum_n \int\frac{d^3p}{(2\pi)^3}tr[G_0^{+}], \ \ 
<{\bar q}_{1} q^{1}> = -iT \sum_n \int\frac{d^3p}{(2\pi)^3}tr[G^{+}].
\end{eqnarray}

The parameters $m_0=0.0055 {\rm GeV}, G_S=5.320 {\rm GeV}^{-2}, 
\Lambda_f=0.637 {\rm GeV}$ are determined by fitting meson properties, 
the parameter $G_D$ in principle can be determined by the nucleon properties, 
here we choose $G_D=2/3 G_S$. 

Our numerical calculations are at $T=0$. In the case without diquark 
condensate, we solve the chiral gap equation as a function of chemical 
potential $\mu$, and get the corresponding
$\Omega(\sigma,\Delta=0,\mu)$; In the chiral condensate and diquark 
condensate coexisting case, we solve the coupled gap equations Eq. (\ref{gap}), 
and get the corresponding value of $\Omega(\sigma,\Delta,\mu)$. 
The thermodynamic potential $\Omega(\sigma,\Delta,\mu)$
(the triangle points) and $\Omega(\sigma,\Delta=0,\mu)$ (the solid circle points) 
as a function of the chemical potential $\mu$ are shown in Fig.1a, and the
phase transition region is seperated in Fig.1b.  It is found that in both cases, 
there exists a metastable region, where thermodynamic 
potential has two values. The lower thermodynamic potential corresponds to the
stable phase. In the case of no diquark condensate, the critical chemical potential of 
the chiral phase transition $\mu_c^{\chi}=0.3453 GeV$, which is a little larger than that
$\mu_c^{D}=0.3401 GeV$ of the phase transition of color superconductivity. 
It is also found that after the phase transition, the thermodynamic potential 
$\Omega(\sigma,\Delta,\mu) <  \Omega(\sigma,\Delta=0,\mu)$ at the same $\mu$,

In the color superconductivity phase, the different colored quarks' chiral condensates 
are shown in Fig.2a as a function of the chemical potential $\mu$, the
region after the phase transition are plotted seperately in Fig.2b, the triangle points 
and the solid circle points are for the quarks participating and not participating in
the diquark condensate. It is found that the difference between different 
colored quarks' chiral condensates is about $1 {\rm MeV}$, which is very small 
comparing with the value of the chiral condensate and diquark condensate in this phase.
It can be regarded that the color symmetry breaking has little influence on different
colored quarks.

\vspace*{-0.5truecm}
\section*{Acknowledgments}
This work was supported in part by China Postdoctoral Science Foundation, 
the NSFC under Grant No. 10105005, 10135030 and 19925519,
and the Major State Basic Research Development Program under 
Contract No. G2000077407.

\vspace*{-2truecm}
\begin{figure}[ht]
\centerline{\epsfxsize=12cm\epsffile{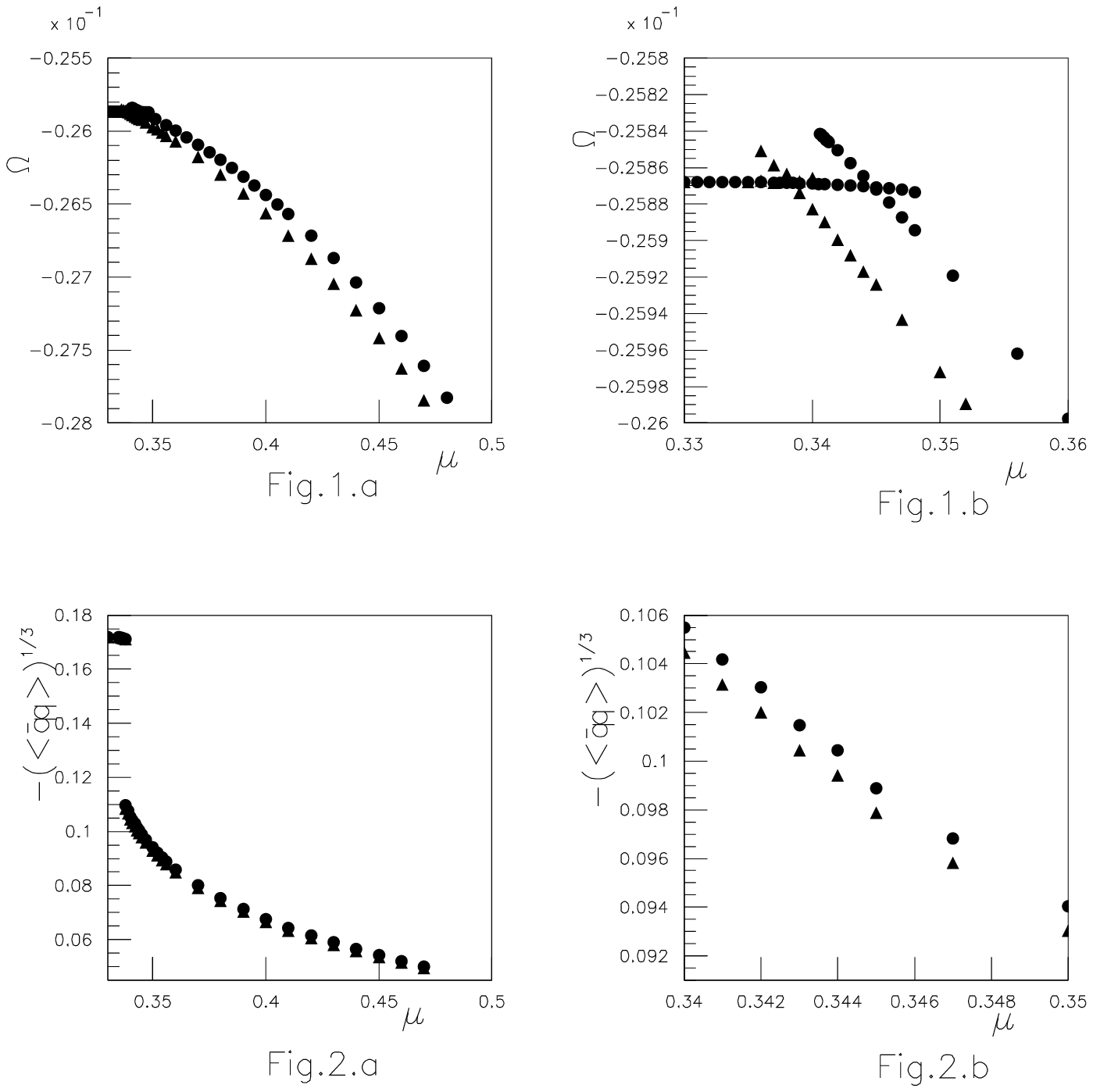}}
\end{figure}

\vspace*{-1truecm}

Fig.1. The thermodynamic potential $\Omega(\sigma,\Delta,\mu)$
(the triangle points) and $\Omega(\sigma,\Delta=0,\mu)$ (the solid circle points) 
as a function of the chemical potential $\mu$ in $a$, and the phase transition region 
is seperated in $b$. \\  

Fig.2. Different colored quarks's chiral condensates 
as a function of $\mu$ in $a$, the region after phase transition is plotted seperately in $b$, 
the triangles and the solid points are for the quarks participating and not participating in
the diquark condensate respectively.

\end{document}